\documentclass[preprint,prd,aps,showpacs,showkeys,nofootinbib]{revtex4}
\usepackage{graphicx}
\begin{document}

\title{SUSY Confinement}

\author{Shu-Min Zhao$^a$\footnote{email:zhaosm@hbu.edu.cn}, Xue-Qian Li$^{b}$\footnote{email:lixq@nankai.edu.cn}}

\affiliation{$^a$Department of Physics, Hebei University, Baoding, 071002, China\\
$^b$Department of Physics, Nankai University, Tianjin, 300071, China}
\date{\today}

\begin{abstract}
In response to the present status that searching for SUSY particles has been unsuccessful,
we propose a bold scenario that SUSY particles are
confined inside hadrons with a required condition of $P_R=1$
in analog to the color confinement for quarks. The scenario seems to be able to reconcile the beautiful
SUSY theory and non-observation at present experiments. On other aspects, some loopholes in the
proposal emerge and require to be answered in the future research.
\end{abstract}

\keywords{SUSY, Confinement}

\maketitle

\section{Introduction}

Following the lesson we learned from the quark confinement we propose a new possible rule which forbids
supersymmetric particles to exist as free ones.
Many years ago, as GellMann established his SU(3) quark model which is so successful to reproduce the "zoo"
of pseudoscalar and vector mesons as well as baryons, the hadron structure became understandable.
However, the quarks carry fractional charges which have
never been observed in all experiments. People tried their best to look for fractional charges because they are
the unique characteristics of free quarks. But all the efforts were ended with failure. On one aspect,
the quark model was so beautiful and successfully explained all hadron structures (including prediction of
$\Omega^-$ ), whereas on the other side, free quarks as the fundamental bricks of the model,  were not experimentally observed.
In a common sense, it exerted
a serious challenge to the theorists of particle physics at that time, namely one must find a proper way out to remedy the severe
loophole.
The answer is the quark confinement, namely a free quark does not exist as an individual object, but
must be combined with other quarks to reside inside mesons or baryons\cite{Luominxing}. Moreover, to satisfy the Pauli exclusion
principle, a new quantum number "color" was introduced. Just as the fractional charges which have never been experimentally
observed, the color quantum number has also not been directly detected in any experiment. Thus
it is natural to associate the quark confinement with a mandatory
condition that if a hadron can exist as a free particle in the nature, it must be in a color-singlet.
To manifest the fact of non-observation of fractional charge, a rule is set which is the consequence of color confinement governed
by non-perturbative QCD.

Since the theory of supersymmetry (SUSY)\cite{supersymmetric models1,supersymmetric models2,supersymmetric models3}
can perfectly solve the hierarchy problem of the Higgs boson and answer many
phenomenological questions, such as the grand unification scale\cite{grand unified1,grand unified2}, the anomalous $g-2$ of muon\cite{g1,g2} etc. one should have
confidence on validity of SUSY. Especially the SUSY theory predicts that all standard model (SM)\cite{SM1,SM2} particles should have
their supersymmetric partners whose spins deviate from their SM partners by 1/2, moreover, the theory also demands the second
Higgs doublet\cite{Higgs1,Higgs2}, so that there must be charged Higgs bosons existing. To signify the difference between SUSY particles and SM
ones, a quantum number R-parity\cite{R-parity1,R-parity2} is introduced as
\begin{equation}
P_R=(-1)^{3(B-L)+2s},
\end{equation}
where $B, L$ and $s$ stand as the baryon number, lepton number and spin of a particle. For a SM particle, $P_R=1$, whereas
for a SUSY particle $P_R=-1$.

After several decades of unsuccessfully searching for SUSY particles,
all experimental results seem to be negative. Does the failure mean that the SUSY theory is not valid or its energy scale
is much higher than people can reach nowadays? Both of them seem not to be natural. In analog to the color
confinement for hadrons, we would propose a rule that the SUSY particles must be confined in hadrons.
It is simple to be realized as long as we demand that $P_R$ must be $+1$ for any particle which can independently
exist as a free particle (on its mass shell) in the nature in analog to the color-singlet requirement.

In this scheme,  not only SM quark, lepton, Higgs boson and gauge bosons physically correspond to their partners squark, slepton,
Higgsino, gauginos, but also the SM composites (SM hadrons) correspond to the SUSY composites (we name them as the SUSY hadrons).
As a matter of fact, a SUSY composite
behaves exactly as a SM composite (meson or baryon or even positronium), at least it is hard
to distinguish them experimentally. By the law, a real particle must be a color singlet
with $P_R=+1$. Namely, as the SUSY ingredients are hidden inside hadrons (mesons or baryons),
only by measuring their spectra, we cannot tell if it is  a SM hadron or a SUSY one.
But possibly, a careful study on their production and decay modes including momentum and angular distributions
of final products (hadrons or leptons), one may expect to gain some information about the parent hadrons (SM or SUSY).

1. The SUSY mesons

Here we still name the particles which contain two SUSY particles as  SUSY mesons, even though they look not
different from SM mesons which we are very familiar with. We write them as $\tilde M$.
They are: (1) $(\tilde q_i\bar{\tilde q'}^i)$ whose spin must be 0 (for ground states, i.e. $l=0$) and charge $0, \pm 1$,
$\tilde q$ is a squark and $\bar{\tilde q}$ is an anti-squark, $i$ is the color number;

(2) $(\tilde l \bar{\tilde l'})$ where $\tilde l$ is a slepton and $\bar{\tilde  l}$ is an anti-slepton and its spin can only be
0 (for ground state), charge can be 0 or $\pm 1$;

(3) $(\tilde{\chi} \bar{\tilde{\chi}'})$ where $\tilde{\chi}$ can be gauginos (photino, gluino, W-ino, Z-ino) and Higgsino, its spin
is 0 or 1, charged or neutral.

Indeed, the stoponium\cite{stoponium} has been studied by several theoretical groups, but the relevant interaction between stop and anti-stop
and the coupling constants may still be somehow of certain arbitrariness.

2. The SUSY baryons

To satisfy the criterion set above, i.e. color-singlet and $P_R=1$, the simplest possible structure for a SUSY baryon ( ground state) is
$N\epsilon_{ijk}q_i\tilde q'_j\tilde q"_k$ where $N$ is the normalization factor, $q_i$ is a SM quark and $\tilde q$
is a super-scalar-quark (squark).  We name such baryons as $\tilde B$ and its spin can only be 1/2  (for the ground state i.e.
the orbital angular momentum is $l=0$).

Definitely, if we allow more ingredients existing in a SUSY baryons, there would be much more candidates, just like
the SM pentaquarks.

3. Possible SU(3) or SU(4) structures.

The SUSY mesons might be arranged according to flavor SU(3) or SU(4) groups, then we can expect new octets and/or
other group representations appearing.

4. The interaction

Charged squarks, sleptons and Wino, charged Higgsinos can interact with each other via exchanging photons, while
colored SUSY particles can also interact via exchanging gluons. But the interaction between SUSY particle and SM one
is still not clear yet.

If we consider the $P_R$ as a conserved quantum number in analog to the color, we may introduce a local symmetry
$U'(1)$ which would lead to existence of an extra gauge boson $Z'$\cite{Z' model1,Z' model2}. As the symmetry is broken, $Z'$ gains
a large mass and would be
a heavy vector boson.
The phenomenology of a new $Z'$-vector boson has
been widely discussed as a possible extension of SM. As a beyond Standard Model (BSM) it has been applied to
study the phenomena observed at high energy experiments, especially it serves as an intermediate agent to bridge between
the dark matter\cite{dark matter1} (what ever it is) and SM particles in detectors on the earth.

Moreover, for the confinement potential, one may still use the phenomenological models where a linear potential
exists to confine ingredients. But it might be different from the QCD confinement potential, i.e. even though the form is similar
to the QCD confinement, there should be an extra factor related to $P_R$, such as $(1-P_R)$, to be responsible for confining two SUSY particles
into a SUSY hadron, meanwhile ruling out the hadron structure of $\bar q_i\tilde q^i$ at all.

By the assumption, one can calculate the binding energies and masses of the SUSY mesons in terms of the Schr\"odinger equation
or other more sophisticated methods.  However since so far we do not have any information about the masses of bare SUSY particles and
interaction coupling constants, we cannot give a quantitative estimate. We will make a bit more discussions on this issue below.

5. Candidate of dark matter (DM).

Besides the electric charges, let us assume that the U'(1) charge of spartilce is opposite to that
of anti-sparticle (could be squark, slepton, chargino, and even neutralinos etc.).
Now most of research tends to consider the weakly interacting heavy particles (WIMPs) as favorable candidates of dark matter. The criterion
of being a dark matter particle is that it does not interact with electromagnetic field and has a long lifetime. The majority
of theorists believe that the most plausible WIMP (weakly interacting massive particle), dark matter candidate, is the neutralino\cite{neutralino}
which is the lightest eigenstate of the mixing of
photino, zino and two CP-even Higgsinos.
However, after several decades of world-wide arduous search on earth and in space, no significant
signals were observed. The direct detection on DM is based on the belief that the WIMPs would interact (maybe weakly) with the SM particles at our
earth detectors. Now the newly measured upper bound  is  $8\times 10^{-42}\; cm*2$ and $3\times 10^{-36}\; cm^2$
for spin-independent and spin-dependent cross sections of WIMPs-nucleon interaction\cite{CDEX} respectively. That value is
already close to the neutrino floor, thus one would ask if the WIMPs indeed
come or if they interact with SM particles via weak interaction?

On the other aspect, there are indirect detections of DM, namely, assuming the dark matter particles would annihilate or decay into some
observable SM particles. So far, some anomalous events, for example, the $e^+$ excess observed in spectra of high energy cosmic rays
indicate that the DM  processes might be the source of the anomalies.

In our scheme, the SUSY baryon cannot be the DM particle because it contains a SM quark which is charged. Moreover,  $\tilde q-\bar{\tilde q}$
composed with charged squark-antisquarks
cannot be a dark matter candidate either. Even though it is neutral, it may have electric and magnetic dipoles, so interacts with electromagnetic
fields. Neither the SUSY meson composed of charged gauginos or Higgsinos due to the same reason.
The proposed dark matter particle can only be a SUSY meson which is composed of a
pair of neutralino-anti-neutralino $\tilde M=\tilde\chi\bar{\tilde\chi}$. The neutralino
interacts with the $Z'-$boson which
may also interact with the SM particles via a small mixing between $Z'$ and photon. The scenario of mixing between $Z'$ and $\gamma$ has been intensively
discussed in literature.

However, as aforementioned, the $U'(1)$ charge of a SUSY particle is opposite to that of its anti-partner, so that when one tries to directly
detect such a SUSY meson $\tilde M$
at the earth detector, its two components ($\tilde \chi$ and $\bar{\tilde\chi}$) destructively contribute to the
scattering processes off the SM particles. Thus the effective interaction between $\tilde M$ and SM particles is remarkably suppressed.
Probably, that is the reason why so far, no signal of DM has been caught by the earth detector.

6. The indirect detection of DM

Even though the probability of direct detection of DM is suppressed, the two components in $\tilde M$ can annihilate into $Z'$ which later turns
into SM particles via mixing with photon, such $e^+e^-$ or $\mu^+\mu^-$ pairs which can be observed by the concerned satellite observations.

Moreover, it may be produced at the high energy colliders.

7. A possibility

As we proposed, the SUSY mesons and baryons are made of the SUSY particles with a mandatory condition of $P_R=1$, so that
they behave just like any ordinary mesons and baryons which are composed of SM quarks and anti-quarks or the positronium composed
of $e^+e^-$ and do not need to be too heavy. If so, one may conjecture that such SUSY hadrons already exist
and have been observed, but have not been properly identified. At present, there are many new particles which cannot be attributed into
the ordinary quark structure and named as X, Y, Z particles, in fact, some of them might indeed be the SUSY hadrons.

Since all the experimentally observed final states only contain hadrons and leptons, it needs an effective way to identify
their ingredients, and this procedure would depend on a smart theory.

8. Probability of producing such SUSY mesons and baryons.

Besides the cosmological processes, we hope to produce the SUSY hadrons through high energy colliders. It is possible
to produce the SUSY hadrons
as long as the relevant scale is not too high. In our ansatz, no very high energy scale is required and probably such hadrons
already exist and wait to be identified. Thus  LHC, especially the proposed CEPC and SPPC\cite{CEPC} would play an important role
to produce the SUSY hadrons. But it is easy to conjecture, the theoretical work is crucially needed to help careful analysis
of data and find characteristics of such SUSY hadrons which should be distinguished from the SM hadrons.

\section{Conclusion}

In this letter we propose a rule that in analogy to the color confinement, the SUSY
particles must be confined inside hadrons or positronium to keep
$P_R=1$. If the mandatory condition is respected, many puzzles seem to find a reasonable
solution.

However, the allegation may also raise several serious problems to be answered by the
future explicit research (both theoretical and experimental). Now let us list some
unavoidable problems.
\begin{enumerate}
\item  What force binds the  SUSY ingredients together into $\tilde{\bar q}\tilde q$,
$\tilde{\chi}^\dagger\tilde\chi$, $\tilde {\bar l}\tilde l$ and $q\tilde q\tilde q$. There may exist a
gauge field similar to $SU_c(3)$ QCD which binds the colored-ingredients to constitute color singlet hadrons.
But definitely a non-perturbative effect in analog to QCD which result in a linear-type confinement for quarks, is needed. Thus,
it is hard, so far, to establish a suitable dynamical framework which can be responsible for SUSY hadrons.

\item According to an intuition, we wonder if all SUSY particles must be too heavy, since it is not
quite natural and necessary from a point of the SUSY theory itself.
However so far nobody knows what is the energy scale for SUSY. We further guess that some of the
newly discovered X,Y,Z particles which cannot be attributed to the regular hadrons might be such SUSY bound states.

\item As we assumed, the direct interaction of the SUSY composites with the SM particles are suppressed, so
$\tilde{\Theta}=\tilde{\bar \chi}\tilde \chi$ as the WIMP DM candidate would be hard to be detected by the earth detectors. However,
the two SUSY ingredients can annihilate into SM particles, such as lepton pairs which result in an excess
of positrons in cosmic rays. But if the annihilation is indeed the cause of the excess, such interaction
might induce decays of $\tilde{\Theta}$ and might contradict to the necessary condition that the DM
particles must be sufficiently stable, at least its lifetime is required to be longer than the lifetime of the present universe.
This requirement constrains the coupling of $\tilde{\bar \chi}\tilde \chi Z'$ where $Z'$ may be a new heavy gauge
boson and as well the mixing parameter of $Z'$ with the SM $\gamma$ photon or $Z$ boson.

\end{enumerate}

All the above questions are waiting for answers, but since this scenario can give reasonable interpretations to some standing
puzzles in the present theory, such as the hierarchy of Higgs and other experimental data, it is still a vital possibility.
Surely, the future theoretical studies and experimental efforts would jointly prove the validity of the scenario.
Of course, if the future experiment can identify a free SUSY particle, it would declare the death of
our proposal after all.

\begin{acknowledgments}

This work is supported by the National Natural Science Foundation
of China (NNSFC) under contract No. 11375128,  11675082, 11735010, 11535002.

\end{acknowledgments}

%%%%%%%%%%%%%%%%%%%%%%%%%%%%%%%%%%%%%%%%%%%%%%%%%%%%%%%%%%%%%%%%%%%
\end{document}